\documentclass[10pt,a4paper]{article}
\pdfoutput=1
\usepackage{changepage,comment}
\usepackage[utf8]{inputenc}
\usepackage{textcomp,marvosym}
\usepackage[]{subfigure}
\usepackage{fixltx2e}
\usepackage{amsmath,amssymb}
\usepackage{pdflscape}
\usepackage{afterpage}
\usepackage{capt-of}
\usepackage{cite}
\usepackage{nameref,hyperref}
\usepackage[right]{lineno}
\usepackage{microtype}
\DisableLigatures[f]{encoding = *, family = * }
\usepackage{rotating}
\usepackage[usenames,dvipsnames,svgnames,table]{xcolor}
\usepackage{colortbl}
\usepackage{epstopdf}
\AppendGraphicsExtensions{.tif}
\usepackage[aboveskip=1pt,labelfont=bf,labelsep=period,justification=raggedright,singlelinecheck=off]{caption}

\makeatletter
\renewcommand{\@biblabel}[1]{\quad#1.}
\makeatother

\date{}

\usepackage{pbox}
\usepackage{pdfpages}

\newcommand{\ac}{\textrm{ac}}
\newcommand{\DW}{\textrm{DW}}
\newcommand{\ATP}{\textrm{ATP}}
\newcommand{\ATPM}{\textrm{ATPM}}
\newcommand{\cat}{\textrm{cat}}
\newcommand{\eq}{\textrm{eq}}

\newcommand{\red}[1]{\textcolor{black}{#1}}

\begin{document}

\begin{flushleft}
{\LARGE
\textbf{\textsf{Constrained Allocation Flux Balance Analysis}
}}
\newline
\\
Matteo Mori\textsuperscript{1,2,3},
Terence Hwa\textsuperscript{3,4},
Olivier C. Martin\textsuperscript{5},
Andrea De Martino\textsuperscript{1,6,7,8\ddag\P},
Enzo Marinari\textsuperscript{1,6,9 \ddag}
\\
\bigskip
{$^1$} Dipartimento di Fisica, Sapienza Universit\`a di Roma, Rome, Italy,
{$^2$} Departamento  de  Bioqu\'imica y Biolog\'ia  Molecular  I,  Universidad  Complutense de  Madrid, Madrid, Spain,
{$^3$} Department of Physics, University of California at San Diego, La Jolla, California, USA,
{$^4$} Institute for Theoretical Studies, ETH Zurich, Switzerland,
{$^5$} GQE - Le Moulon, INRA, Univ. Paris-Sud, CNRS, AgroParisTech, Universit\'e Paris-Saclay, Gif-sur-Yvette, France,
{$^6$} Soft and Living Matter Lab, Istituto di Nanotecnologia (CNR-NANOTEC), Consiglio Nazionale delle Ricerche, Rome, Italy,
{$^7$} Center for Life Nano Science@Sapienza, Istituto Italiano di Tecnologia, Rome, Italy,
{$^8$} Human Genetics Foundation, Turin, Italy,
{$^9$} INFN, Sezione di Roma 1, Rome, Italy

\bigskip

\ddag Authors contributed equally to this work\\
\P andrea.demartino@roma1.infn.it

\end{flushleft}

\section*{\textsf{Abstract}}

New experimental results on bacterial growth inspire a novel top-down approach to study cell metabolism, combining mass balance and proteomic constraints to extend and complement Flux Balance Analysis. We introduce here Constrained Allocation Flux Balance Analysis, CAFBA, in which the biosynthetic costs associated to growth are accounted for in an effective way through a single additional genome-wide constraint. Its roots lie in the experimentally observed pattern of proteome allocation for metabolic functions, allowing to bridge regulation and metabolism in a transparent way under the principle of growth-rate maximization. We provide a simple method to solve CAFBA efficiently and propose an ``ensemble averaging'' procedure to account for unknown protein costs. Applying this approach to modeling {\it {\it E. coli}} metabolism, we find that, as the growth rate increases, CAFBA solutions cross over from respiratory, growth-yield maximizing states (preferred at slow growth) to fermentative states with carbon overflow (preferred at fast growth). In addition, CAFBA allows for quantitatively accurate predictions on the rate of acetate excretion and growth yield based on only 3 parameters determined by empirical growth laws. 


~

\section*{\textsf{Author Summary}}

The intracellular protein levels of exponentially growing bacteria are known to vary strongly with growth conditions, as described by quantitative ``growth laws''. This work introduces a computational genome-scale framework (Constrained Allocation Flux Balance Analysis, CAFBA) which incorporates growth laws into canonical Flux Balance Analysis. Upon introducing 3 parameters based on established growth laws for {\it {\it E. coli}}, CAFBA accurately reproduces empirical results on the growth-rate dependent rate of carbon overflow and growth yield, and generates testable predictions about cellular energetic strategies and protein expression levels. CAFBA therefore provides a simple, quantitative approach to balancing the trade-off between growth and its associated biosynthetic costs at genome-scale, without the burden of tuning many inaccessible parameters.

\section*{\textsf{Introduction}}

The coupling between the physiology of cell growth and cellular composition has been actively investigated since the 1940s. In exponentially growing bacteria, whose growth state is conveniently associated to a single parameter, namely their growth rate, such interdependence is best expressed in a quantitative way by the  bacterial `growth laws' that directly relate the protein, DNA and RNA content of a cell to the growth rate. Many such laws have been experimentally characterized \cite{1,2,3,4} and many more are currently being probed at increasingly high resolution \cite{5,6}. The emerging scenario suggests that proteome organization in bacteria is actively regulated in response to the growth conditions. Recent experiments have in particular provided validation to the picture according to which, as the growth rate changes, bacteria adjust the relative amounts of ribosome-affiliated, nutrient scavenging and metabolic proteins (enzymes), so as to optimize their growth performance and energy production strategy \cite{6,7,8}. At present, several phenomenological models explain the origin of different growth laws at a coarse-grained level \cite{5,7}. In contrast, genome-scale approaches probing such relationships at molecular levels are less developed.

Constraint-based models (CBMs) are powerful in silico tools that can be used to examine metabolic networks at genome scale. Starting from a non-equilibrium steady state assumption for metabolic fluxes, CBMs define the space of feasible reaction profiles through simple physico-chemical constraints like mass-balance. Once physiologically or thermodynamically motivated bounds of variability are assigned to fluxes, the solution space is essentially determined by the stoichiometry of the network alone. On the other hand, in genome-scale models stoichiometric constraints usually generate high-dimensional solution spaces in which physiologically relevant flux patterns may be hard to isolate. In many cases, optimal flux patterns can be defined through the maximization of specific objective functions. Flux Balance Analysis (FBA) \cite{9,10,11,12,13,14,15} allows for instance to compute optimal flux configurations by means of linear programming (LP), employing biomass production as a standard objective function \cite{16}. This approach is widely used to describe microbial growth in lab conditions. 

It is clear that in order to capture the phenomenology of growth laws one needs to go beyond the basic elements of CBMs, and incorporate the costs associated with gene expression and protein synthesis into models of cellular metabolism. Resource Balance Analysis (RBA) \cite{17,18} and ME-models \cite{19,20} have taken important steps in this direction. These approaches propose a data-based optimization scheme to predict the growth-maximizing metabolic flux configurations under a variety of constraints, including stoichiometric mass-balance, `demand functions' characterizing how the amounts of cellular components change with the growth rate, and specific prescriptions that relate fluxes to enzyme levels. The resulting schemes are more involved than FBA (resulting in nonlinear optimization problems) and require a large number of parameters. It is therefore important to devise a theoretical framework with the conceptual appeal and computational simplicity of FBA, in particular one that is more resilient to the choice of parameters and in which the interplay between metabolism and regulation is expressed through a more intuitive and transparent framework.

In this work we present a generalized FBA scheme, called Constrained Allocation FBA or CAFBA, in which (optimal) regulation is accounted for effectively through a single additional global constraint on fluxes that encodes for the relative adjustment of proteome sectors at different growth rates. In a nutshell, the CAFBA-specific constraint describes the tug-of-war in the allocation of cellular resources across ribosomal, transport and biosynthetic proteins that has been observed in experiments. By imposing that the ribosomal share of the proteome behaves in accordance with empirically established growth laws \cite{5,21, 22}, CAFBA is able to reproduce observed behaviors without requiring parameter tuning. In addition, CAFBA generates a variety of testable predictions, including about the usage of metabolic pathways, despite lacking the level of biochemical detail that characterizes ME-models or RBA. 

Cellular strategies for energy production are the central focus of CAFBA. It is well known that fast-growing microorganisms tend to avoid using high-yield respiratory pathways to generate ATP even in the presence of oxygen, relying instead on aerobic fermentation \cite{23,24,25,26,27,28}. The preference for low-yield pathways is manifested in the secretion of fermentation products like acetate for {\it {\it E. coli}} or ethanol for {\it S. cerevisi\ae} \cite{23,25,26,29}. This phenomenon, known as `overflow metabolism', is captured by standard FBA schemes at a qualitative level when additional capacity constraints on respiratory pathways \cite{30} or density constraints for soluble \cite{31, 32} or membrane-bound \cite{33} enzymes are included. However, certain quantitative aspects of potential interest for industrial applications, like the rate of metabolic overflow and the growth rate at which it occurs, have so far eluded comprehensive mechanistic models. By effectively modeling the trade-off between growth and its biosynthetic costs, CAFBA naturally produces cellular states with suboptimal growth yields, where carbon overflow is obtained with quantitative accuracy.

This paper focuses on the scenario obtained by CAFBA for carbon-limited growth of {\it {\it E. coli}}. We find in particular that acetate secretion appears in {\it {\it E. coli}} at fast growth rates, whereas yield-maximizing FBA-like solutions dominate at slow growth rates. In spite of the nominal need for a large number of uncharacterized parameters in genome-wide models, CAFBA solutions remarkably depend only on a few global parameters. In particular, overflow metabolism is obtained consistently with quantitative accuracy, while all results are robust against 10-fold changes in the values of the enzymatic efficiency parameters. From a technical viewpoint, CAFBA effectively turns out to be an LP problem even when one accounts for growth-rate dependent biomass composition. This, together with its simple conceptual framework, makes CAFBA a very convenient scheme to analyze the interplay of metabolism and gene expression at genome scale.

\section*{\textsf{Model}}

\subsection*{\textsf{Proteome sectors}}

Phenomenological studies of bacterial growth physiology suggest that the bacterial proteome is organized into ``sectors'' whose mass fractions adjust linearly with the growth rate in response to specific environmental and intracellular changes, including carbon limitation, anabolic limitation and translational inhibition \cite{5,6,8}. Proteome organization and optimal growth constitute in essence an intertwined allocation problem, with the cell trying to optimally partition its proteome so as to maximize its growth performance. Based on empirical evidence on {\it {\it E. coli}} growth in carbon-limited media, CAFBA posits a 4-sector partitioning of the proteome in
\begin{itemize}
\item ribosome-affiliated proteins (R-sector);
\item biosynthetic enzymes (E-sector);
\item proteins devoted to carbon intake and transport (C-sector); \item core housekeeping proteins whose expression level is independent of the growth rate (Q-sector).
\end{itemize}
The corresponding proteome fractions (denoted by $\phi_X$ for the X-sector) should sum up to 1, i.e.
\begin{equation}\label{one}
\phi_C+\phi_E+\phi_R+\phi_Q=1~~.
\end{equation}
We shall now provide an explicit characterization of the different terms in the above sum.

\paragraph*{\textsf{The ribosomal sector.}} $\phi_R$ is experimentally found to be linearly dependent on the growth rate $\lambda$ when growth is nutrient-limited \cite{3,4,5}, namely
\begin{equation}\label{two}
\phi_R=\phi_{R,0}+w_R\lambda~~,
\end{equation}
where $\phi_{R,0}$ is a strain-dependent constant representing the extrapolated ribosomal proteome fraction at zero growth rate, and $w_R$ is a strain-independent constant related to the ribosome's translational efficiency \cite{5,6}. Phenomenologically, $w_R$ describes the proteome fraction allocated to ribosomal proteins per unit of growth rate. At the molecular level, the linear relation (\ref{two}) is enforced by a regulatory mechanism involving the alarmone ppGpp \cite{34,35,36,37}. When focusing on carbon-limited growth, one can set $w_R$ equal to the empirical value $w_{R,0}\simeq 0.169$ h \cite{5}. The effects of translational inhibition can instead be studied by increasing $w_R$ from the value $w_{R,0}$, so as to model the increasingly slowed-down translation induced by antibiotics \cite{22}. As will soon become clear, the value of the offset $\phi_{R,0}$ is immaterial for the formulation of CAFBA.

\paragraph*{\textsf{The carbon catabolic sector.}} We focus on balanced growth in a minimal medium containing a single carbon source (e.g. glucose). Based on experimental findings \cite{6,8}, we assume that $\phi_C$ depends linearly on the carbon intake flux $v_C$, i.e. 
\begin{equation}\label{three}
\phi_C=\phi_{C,0}+w_C v_C~~,
\end{equation}
where, by analogy with (\ref{two}), $\phi_{C,0}$ is a $\lambda$-independent offset and $w_C$ characterizes the proteome fraction allocated to the C-sector per unit of carbon influx. Recent proteomic studies \cite{8,38} suggest that the C-sector should include not only the specific transport system taking up the sugar, but also other proteins that are co-expressed in response to carbon limitation through mediation by the pleiotropic regulator cAMP-Crp \cite{6}, like intake proteins for other nutrients, motility proteins, etc. Therefore, (\ref{three}) should be seen as an effective prescription accounting for the fact that several types of proteins intended for nutrient scavenging and intake are co-expressed in carbon limitation. All of these should be expected to contribute to $\phi_C$, even if certain proteins, like motility proteins, may not be required for growth in laboratory conditions. The offset $\phi_{C,0}$ thus represents a basal level of proteins  not due to carbon intake only.

In order to better characterize $w_C$ (i.e. the carbon-intake dependent part of $\phi_C$), we assume that the carbon influx $v_C$ at a given extracellular sugar level $[g]$ is described by a Michaelis-Menten kinetics of the form
\begin{equation}\label{four}
v_C=\frac{V}{M_{\DW}}k_{\cat,g}[E_g]\frac{[g]}{[g]+K_{M,g}}~~,
\end{equation}
where $[E_g]$ stands for the level of the intake protein(s) specific to $g$ that are not in $\phi_{C,0}$, $k_{\cat,g}$ and $K_{M,g}$ are kinetic constants, and $V$ and $M_{\DW}$ represent the cell volume and dry weight, respectively. (The ratio $V/M_{\DW}$ is introduced so that the flux units are mmol$/$g$_{\DW}$h.) Denoting the total protein mass by $M_{TP}$ and the enzyme's molecular weight by $\mu_g$, and letting $\alpha_g$ be the mass fraction of enzyme $E_g$ in the C-sector, we can express $[E_g]$ in terms of the C-sector's proteome  fraction $\phi_C$ as $[E_g]=(\phi_C-\phi_{C,0})\alpha_g M_{TP}/(\mu_g V)$.  In turn, (\ref{four}) can be rewritten as 
\begin{equation}\label{five}
v_C=\kappa_{\cat,g}(\phi_C-\phi_{C,0})\frac{[g]}{[g]+K_{M,g}}~~,
\end{equation}
where $\kappa_{\cat,g}\equiv \alpha_g(k_{\cat,g}/\mu_g)(M_{TP}/M_{\DW})$. The factor $M_{TP}/M_{\DW}\simeq 60\%$ is roughly constant for a wide range of growth rates \cite{39, 40}, $k_{\cat,g}/\mu_g$ is instead an enzyme-specific property,  while the proportion $\alpha_g$ is determined genetically by the expression level of the enzyme $E_g$ relative to those of the other co-expressed C-sector proteins. Comparing (\ref{five}) to  (\ref{three}), one sees that $w_C$ can be represented as \begin{equation}\label{six}
w_C=w_{C,0}\left(1+\frac{K_{M,g}}{[g]}\right)~~,
\end{equation}
with $w_{C,0}\equiv 1/\kappa_{\cat,g}$. 

The above analysis suggests that $w_C$ can be conveniently used to control the carbon influx: it takes on a sugar-specific value $w_{C,0}$ at saturating sugar concentrations (i.e. for $[g]\gg K_{M,g}$) and the effect of reducing extracellular sugar levels can be modeled by simply increasing its value. Hence, as a proxy of varying the abundance of the carbon source, we will simply dial $w_C$. The importance of using $w_C$ as control parameter, as opposed to varying the maximum nutrient intake capacity, is discussed in Note B in the Supporting Text. Note that the maximal growth rate achievable in the  medium we consider (referred to as $\lambda_{\max}$ below, and obtained for $w_C=0$ or, equivalently, $w_{C,0}=0$) is experimentally determined by the extrapolated growth rate at which C-sector protein expression vanishes \cite{6, 41}. 

\paragraph*{\textsf{The biosynthesis sector.}} The flux through enzymatic reactions involved in biosynthesis (E-sector) can be generally written in the form
\begin{equation}\label{seven}
v_i=\frac{V}{M_{\DW}}k_{\cat,i}[E_i] \,f_i([s_i],[p_i])~~,
\end{equation}
where $[E_i]$ denotes the concentration of enzyme $i$ and we considered explicitly an additional dependence on the concentrations of the substrates ($[s_i]$) and products ($[p_i]$) through the function $f_i$. For an elementary irreversible reaction with a single substrate and a single product, $f_i$ is a Michaelis-Menten function of $[s_i]$, while for reactions close to thermodynamic equilibrium $f_i\simeq [s_i]-[p_i]/K_{\eq}$ \cite{42}. In full analogy with the previous case, we can express $[E_i]$ in terms of the proteome fraction $\phi_i$ of enzyme $E_i$ as  $[E_i]=\phi_i M_{TP}/(\mu_i V)$. Defining $\kappa_{\cat}\equiv (k_{\cat,i}/\mu_i)(M_{TP}/M_{\DW})$, we then have 
\begin{equation}\label{eight}
v_i=\kappa_{\cat}\phi_i\,f_i([s_i],[p_i])~~.
\end{equation}
Motivated by the observed linear dependence between enzyme abundance and growth rate in carbon-limited growth \cite{6,8} and assuming the generic linear dependence between biosynthetic flux and growth rate, we set
\begin{equation}\label{nine}
\phi_i=\phi_{i,0}+w_i |v_i|~~,
\end{equation}
with a fixed offset $\phi_{i,0}$. The ``weight'' $w_i$ represents the proteome fraction to be invested in enzyme $E_i$ per unit flux of reaction $i$. The absolute value instead reflects the fact that, for a reversible process, a protein cost has to be faced independently of the net direction. Note that, in principle, the values of the weights $w_i$ can be determined experimentally by fitting, for each reaction, proteomic and flux measurements at different growth rates to (\ref{nine}).

The linear relation (\ref{nine}) can be directly obtained from (\ref{eight}) assuming that reaction $i$ is irreversible and that the enzyme $E_i$ is operating in the saturated regime. In this case, $\phi_{i,0}=0$ and $w_i=1/\kappa_{\cat,i}$. However, such reactions would be incapable of balancing flux in the event of transient changes, leading to the accumulation of intermediate metabolites. Therefore, most intracellular reactions in physiological conditions should not be expected to operate in the saturated regime. Reactions carrying a flux proportional to the substrate level (as in flux sensors \cite{5,43} and charged tRNAs \cite{35}) can again be described by (\ref{nine}), albeit with an offset $\phi_{i,0}$; see Note A in the Supporting Text. In this view, the offset $\phi_{i,0}$ provides a mathematically simple way to capture the fact that, at slow growth, the flux approaches zero due to adjustments in metabolite pools while enzyme levels remain finite. As for the other sectors, the values of these offsets  play no role in CAFBA (see below). Summing up the contributions of each reaction, the proteome fraction of the E-sector $\phi_E\equiv\sum_i\phi_i$ can be written as 
\begin{equation}\label{ten}
\phi_E=\phi_{E,0}+\sum_i w_i|v_i|~~,
\end{equation}
where the sum runs over all enzyme-catalyzed reactions and $\phi_{E,0}\equiv\sum_i \phi_{i,0}$ contributes to a core, $\lambda$-independent proteome fraction for baseline expression levels \cite{6,8}.

\subsection*{\textsf{Proteome-wide constraint}}

Putting the different terms together, the sum rule (\ref{one}) for proteome fractions can be recast as
\begin{equation}\label{eleven}
w_C v_C+\sum_i w_i |v_i|+w_R\lambda=\phi_{\max}~~,
\end{equation}
where $\phi_{\max}=1-\phi_Q-\phi_{C,0}-\phi_{E,0}-\phi_{R,0}$ denotes the proteome fraction accessible to growth-rate dependent components of the protein sectors, which was estimated to be of the order of 50\% for {\it {\it E. coli}} \cite{5}. The linear constraint (\ref{eleven}) encodes for the tug-of-war that ultimately determines optimal growth and proteome allocation, as depicted in Fig 1A: as $\lambda$ increases, so does the proteome fraction of the R-sector, and the E- and C- sectors will concomitantly have to adjust their shares so as to satisfy (\ref{eleven}), forcing in turn a remodeling of the underlying flux and nutrient intake patterns. 

\begin{figure}
\begin{center}
\includegraphics[width=10.5cm]{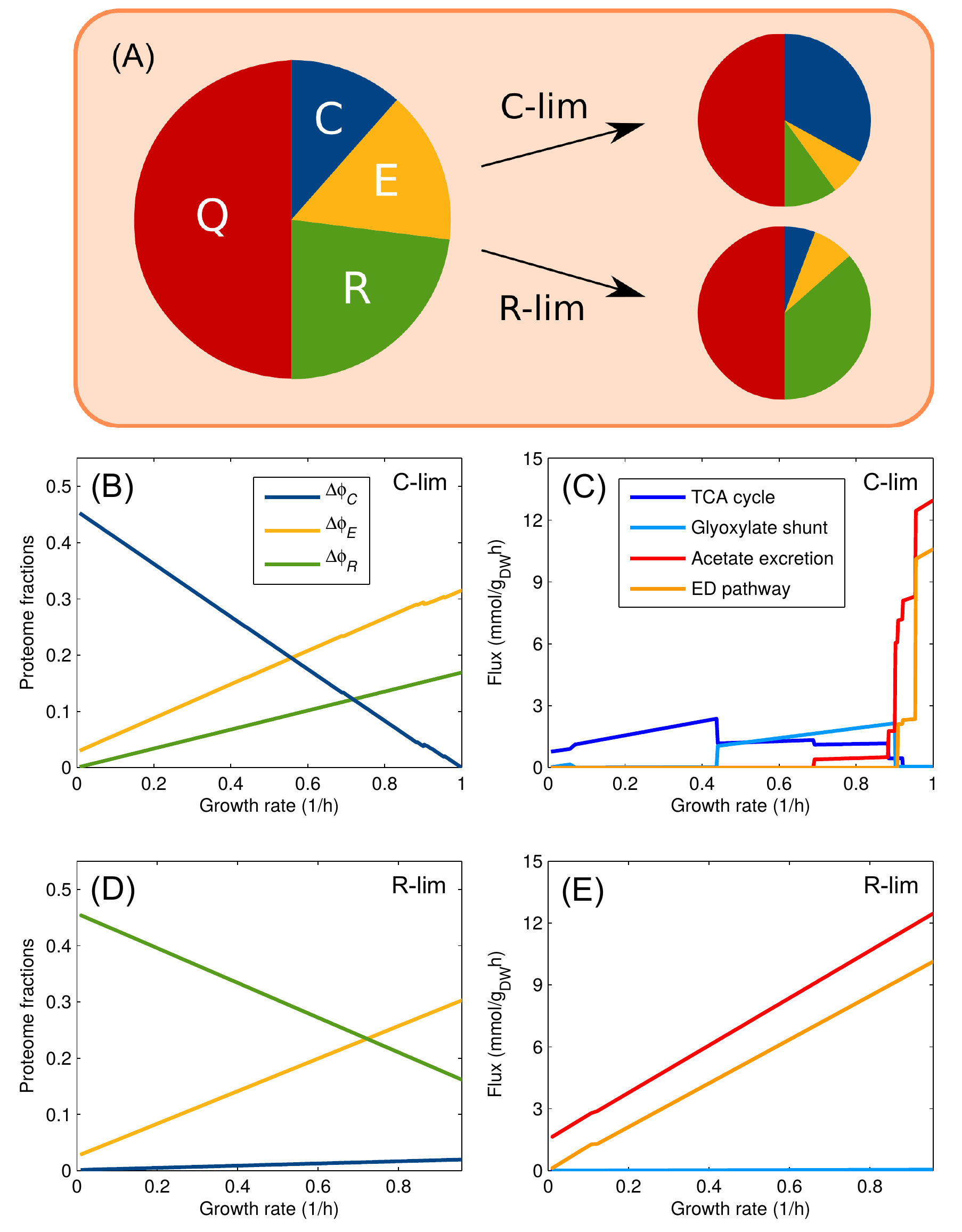}
\end{center}
\caption{\textbf{CAFBA solutions for {\it {\textbf{E. coli}}} in the homogeneous case for carbon limitation and translational limitation.} (A) Proteome organization in CAFBA: R-sector of ribosome-affiliated proteins (growth rate dependent), E-sector of ``enzymes'' (flux-dependent), C-sector of catabolic proteins (dependent on the carbon influx), and a fixed Q-sector of ``housekeeping'' proteins. The fractions in these four sectors sum up to one. C-, E- and R- sectors adjust their size depending on the environmental conditions, while the Q-sector accounts for roughly 50\% of the proteome. We model the three growth-dependent sectors as a constant plus a variable part, i.e. $\phi_X=\phi_{X,0}+\Delta\phi_X$ with $X\in\{C,E,R\}$. $\Delta\phi_C=w_c v_C$ is proportional to the carbon intake flux; $\Delta\phi_E=\sum_i w_i|v_i|$ is a weighted sum of non-catabolic fluxes; $\Delta\phi_R=w_R\lambda$ is proportional to the growth rate $\lambda$. (B) Growth rate-dependent parts of proteome sectors plotted versus $\lambda$ in carbon limitation (C-lim). As the external glucose concentration is reduced, more catabolic proteins are needed per unit of carbon influx. The cell allocates a larger share to C-proteins, while reducing the E- and R-sector shares. (C) CAFBA fluxes as a function of $\lambda$, obtained by varying the degree of carbon (glucose) limitation (C-lim). A transition from fermentation to respiration appears when growth rate is in the range 0.7--0.9/h. The Embden-Doudoroff pathway and the glyoxylate shunt are both operated at high growth rates. (D) The $\lambda$-dependent parts of the proteome sectors plotted against growth rate in translational limitation (R-lim). This is obtained by keeping $w_C$ constant while increasing $w_R$, thereby simulating increasing levels of translation-inhibiting antibiotics. The cell allocates more proteins to the ribosomal sector while reducing the proteome share devoted to carbon metabolism and biosynthesis. (E) CAFBA fluxes as a function of $\lambda$ obtained in R-limitation for increasing values of $w_R$, at constant $w_C$. Acetate is secreted at low growth rates if the  extracellular carbon level is large enough. Fluxes through the TCA cycle, the glyoxylate shunt and the Entner-Doudoroff pathway are represented by $\alpha$KG dehydrogenase, malate synthase and 6-phosphogluconate dehydratase fluxes, respectively. In panels B and C, corresponding to C-limitation, $w_R$ was set to $w_{R,0}=0.169$ h, while R-limitation (panels D and E) was obtained using $w_C=1.4\times 10^{-3}$ gh/mmol, corresponding to a carbon source with high nutritional capacity. In both cases we set all weights in the E-sector to $w_E=8.3\times 10^{-4}$ gh/mmol and $\phi_{\max}=48.4\%$.}
\end{figure}

Formally, the proteome allocation constraint (\ref{eleven}) resembles the molecular crowding constraint defined in \cite{31,32}, which essentially enforces a global upper bound on fluxes due to finite solvent capacity and was also adopted in RBA \cite{17,18}. However, the intracellular density is empirically known to be (roughly) constant across different growth conditions \cite{44}, suggesting  that cells can adapt their volume to accommodate additional metabolites and macromolecules when necessary. In this respect, a hard constraint on solvent capacity is not fully justified. The CAFBA constraint (\ref{one}) is instead derived from the normalization of protein fractions, due ultimately to the limited translational capacity of the ribosomes \cite{5}. Note that the growth rate $\lambda$ is explicitly involved in (\ref{eleven}).

\subsection*{\textsf{Constrained allocation FBA}}

Summing up, CAFBA is defined by the following optimization problem:
\begin{eqnarray}\label{twelve}
\max_{\mathbf{v}}\lambda~~\text{subject to} & \text{(i)} & \sum_i S_{\mu i}v_i=0 ~~~\forall\mu \nonumber\\
& \text{(ii)} & \ell_i\leq v_i\leq u_i~~~\forall i\\
& \text{(iii)} & w_C v_C+\sum_i w_i |v_i|+w_R\lambda=\phi_{\max} \nonumber
\end{eqnarray}
where $S_{\mu i}$ stands for elements of the metabolic network's stoichiometric matrix (with $\mu$ indexing metabolites and $i$ indexing reactions), $\ell_i$ and $u_i$ denote lower and upper bounds for the flux $v_i$, while the value of $\lambda$ is  defined by the flux of the biomass reaction \cite{16}.

\section*{\textsf{Results}}

We have studied CAFBA solutions for the {\it {\it E. coli}} iJR904 GSM/GPR reconstruction \cite{45} assuming growth limited by a single carbon source (glucose). (See Note C in the Supporting Text for details about CAFBA in different growth-limiting conditions.) We started from the case of $\lambda$-independent biomass composition, in which CAFBA can be solved exactly by LP (see Materials and Methods), and then considered the more general case of growth-rate dependent biomass. Throughout this study, we set the parameters $w_R$ and $\phi_{\max}$ to the values $w_{R,0}=0.169$ h and $\phi_{\max}=48.4\%$ found empirically for {\it {\it E. coli}} K-12 MG1655 \cite{5}. A nearly identical $w_{R,0}$ and a slightly smaller $\phi_{\max}$ have been reported in \cite{6, 29} for {\it {\it E. coli}} K-12 strain (NCM3722). (The values of $w_{R,0}$ and $\phi_{\max}$ need not be fine-tuned. In fact, variations in one of these parameters can be compensated by rescaling the weights of reactions in the E-sector. The dependence of the fluxes on $w_{R,0}$ and $\phi_{\max}$ is described in detail in Note D in the Supporting Text.) Carbon limitation is enforced by increasing the value of $w_C$ from its minimum $w_{C,0}$, corresponding to saturating glucose concentrations. (Likewise, translational limitation can be studied by increasing the value of $w_R$ from its minimum, $w_{R,0}$ \cite{22}.) For each choice of $w_C$ and $w_R$ (and of the set of $w_i$'s), we  solve (\ref{twelve}) for the fluxes $v_i$ that maximize the growth rate $\lambda$.

As said above, the weights $w_i$ could in principle be determined by combining proteomic studies \cite{8,37,46} with direct flux measurements taken in the appropriate growth conditions. However, the coverage by mass spectroscopy is still limited, and the accuracy of protein abundance is often no better than 2-fold. Much better estimates of protein abundances have been obtained recently using ribosome profiling \cite{47}, which also provides a near complete coverage of {\it {\it E. coli}} proteins. This method is however much less versatile compared to proteomics and only a few conditions have been probed so far. Given the lack of reliable empirical estimates, we have focused on two limiting situations allowing us to characterize intrinsic properties of CAFBA for {\it {\it E. coli}} in the most transparent manner.
\begin{itemize}
\item The homogeneous case: here, $w_i$'s are uniformly set to the same value, denoted by $w_E$, for each reaction $i$. $w_E$ is chosen so that the fastest growth rate achievable $\lambda_{\max}$, corresponding to $w_C=0$, matches the corresponding empirical value.
\item The heterogeneous case: here, $w_i$'s are taken at random, to reflect one's lack of knowledge of their specific value. More precisely, $w_i$ is drawn from a prescribed probability distribution independently for each reaction $i$. The mean value $\langle w\rangle$ of the distribution essentially plays the role played by $w_E$ in the homogeneous case, in that it sets the average growth rate obtained for $w_C=0$ to $\lambda_{\max}$. Clearly, the quantitative details of CAFBA solutions will depend on the specific values taken by the $w_i$'s. However, as there is no reason to concentrate on a single set of weights, we will focus our analysis on two aspects, namely (i) the ``average'' behaviour obtained by averaging solutions over different realizations of the $w_i$'s, and (ii) the fluctuations of solutions around this average.
\end{itemize}

\subsection*{\textsf{Homogeneous weights and patterns of flux}}

We set $w_E$ so that the extrapolated maximal growth rate in unlimited carbon supply, corresponding to $w_C=0$, is close to the value $1.1-1.2/$h found in \cite{6,41}. In the case of glucose as the sole carbon source, as well as for a number of other glycolytic carbon sources (see Table A in the Supporting Text), the value $w_E=8.3\times 10^{-4}$ gh/mmol turned out to yield $\lambda_{\max}=1/$h. (Slightly larger growth rates are obtained with phosphorylated carbon sources due to the fact that the extra energy carried by these carbon sources allow for a reduced flux in the E-sector.) To capture the effects of changing the glucose level, we simply increased the value of $w_C$ from zero. For the sakes of completeness, the values of $w_C$ leading to empirically observed growth rates for {\it {\it E. coli}} growth on different carbon sources are reported in Table B in the Supporting Text.

Fig 1 reports results obtained for growth on glucose with this choice of parameters, while results for growth on other carbon sources are shown in Fig A in the Supporting Text. One sees in Fig 1B that the growth-dependent fraction of C-proteins ($\Delta\phi_C$, blue line) increases almost linearly with decreasing $\lambda$ as the carbon concentration is limited, in line with the experimentally observed expressions of catabolic proteins \cite{6,8} and PTS activity \cite{20}. Both the proteome fractions of the E- and R-sectors ($\Delta\phi_E$ and $\Delta\phi_R$, yellow and green lines, respectively) instead decrease linearly with growth rate. CAFBA therefore confirms the findings from a coarse grained model of proteome allocation \cite{7}: in the optimal state, the cell invests more and more of its proteomic resources in intake systems as nutrient becomes limiting, while translational machinery and biosynthetic pathways are favored at high growth rates.

Fig 1C displays the main fluxes of central carbon metabolism. The rate of acetate secretion and the flux through the Entner-Doudoroff pathway (red and orange colors, respectively) both drop fast as the growth rate decreases. Respiration, represented by the flux through the TCA cycle (blue color) is the predominant energy-producing pathway at small growth rates, while at high growth rates fermentation is preferred and acetate is secreted. Note that the acetate onset point is within 10\% of the one observed experimentally for NCM3722 \cite{29} roughly independently of the specific carbon source (see Fig A in the Supporting Text) -- a remarkable result given the simplicity of the homogeneity assumption for the $w_i$'s.

Translational limitation \cite{5} is modeled by increasing $w_R$ from the value of $w_{R,0}$ while keeping all other parameters fixed, including $w_C$. In this case (see Fig 1D), the ribosomal proteome fraction ($\Delta\phi_R$) increases as translation is increasingly inhibited, while the other growth-dependent sectors ($\Delta\phi_C$ and $\Delta\phi_E$) shrink almost linearly. Acetate secretion extends to the slowest growth rates in accordance with experimental findings \cite{29}, while the respiratory flux (see Fig 1E) is negligible.

It is interesting to compare CAFBA results with the phenomenological proteome allocation model introduced in \cite{5}, which describes how proteome is allocated in different environments. There, the growth rate was predicted to be a Michaelis-Menten function of the ``nutritional capacity'' $\kappa_n$ and ``translation capacity'' $\kappa_t$, independent phenomenological parameters that can be estimated from empirical growth laws. CAFBA recovers this result within a genome-scale model, with $1/w_C$ playing the role of $\kappa_n$ and $1/w_R$ acting as $\kappa_t$ (see Fig B in the Supporting Text and Note D in the Supporting Text for a detailed discussion).

It transpires from Fig 1C and Fig A in the Supporting Text that the optimal flux configurations in carbon limitation vary discontinuously with the growth rate. This is due to the fact that the control parameter is not a flux (as in standard FBA), but, rather, the weight of the C-sector $w_C$, which, as discussed above, is a proxy for either the external carbon concentration or the amount of glucose intake proteins \cite{6}. Even though $w_C$ is varied continuously, growth-rate maximization can induce large rearrangements of the active pathways in response to small changes of the control parameter. This behavior is ultimately a mathematical feature due to the way in which the optimal solution in constraint-based models like CAFBA changes as one modifies $w_C$.

\subsection*{\textsf{Heterogeneous weights and patterns of average flux}}

For the heterogeneous case, for each value of $w_C$ we generated 1000 models, each with a random set of weights $w_i$ independently drawn from the same probability density 
\begin{equation}\label{thirteen}
p(w)\propto 1/w~~~,~~~ w_{\min}\leq w\leq w_{\max}~~,
\end{equation}
which corresponds to a uniform density for the logarithm of $w$. $p(w)$ is fully determined by its average $\langle w\rangle$ and width $\delta\equiv\log_{10}(w_{\max}/w_{\min})$. We set $\langle w\rangle$ so that the average value of the maximum achievable growth rate $\lambda_{\max}$ (obtained for $w_{C,0}=0$) equals 1/h. This fixes $\langle w\rangle=8.8\times 10^{-4}$ gh/mmol, a value that is remarkably close to $w_E=8.3\times 10^{-4}$ gh/mmol as determined in the homogeneous case. $\delta$ was instead fixed to 1, implying that the weights are assumed to span one order of magnitude (results obtained for different values of $\delta$ are discussed in Note E in the Supporting Text). 

Each set of weights $\{w_i\}$ leads to a corresponding optimal flux pattern, growth rate, acetate secretion rate, etc. The distribution of growth rates obtained from many realizations of the weights is shown in Fig 2A. Note that in spite of the 10-fold variability of the weights, the growth rate remains within a modest range of $\pm 20\%$. The distribution for acetate secretion rates is instead conveyed in  Fig 2B: it is rather heterogeneous, with a marked peak for phenotypes with very low acetate secretion. While individual fluxes can fluctuate  significantly across solutions, average fluxes are strikingly well-behaved. This phenomenon is illustrated in Fig 2C where we show a set of average fluxes plotted against the average growth rate. The average acetate secretion rate (red symbols) has an approximate linear dependence on the growth rate  starting from $\lambda_{\ac}\simeq 0.79/$h. Average fluxes through TCA and the glyoxylate shunt (blue up- and down- triangles) reach their respective maxima close to $\lambda_{\ac}$. Notice that a smooth transition from a predominantly fermentative to a predominantly respiratory mode of energy production clearly emerges, in full agreement with empirical evidence. It is especially remarkable that this scenario does not seem to depend  on the specific choice of $p(w)$. For instance, a log-normal distribution gives qualitatively similar results (see Fig C in the Supporting Text). 

\begin{figure}[h]
\begin{center}
\includegraphics[width=15cm]{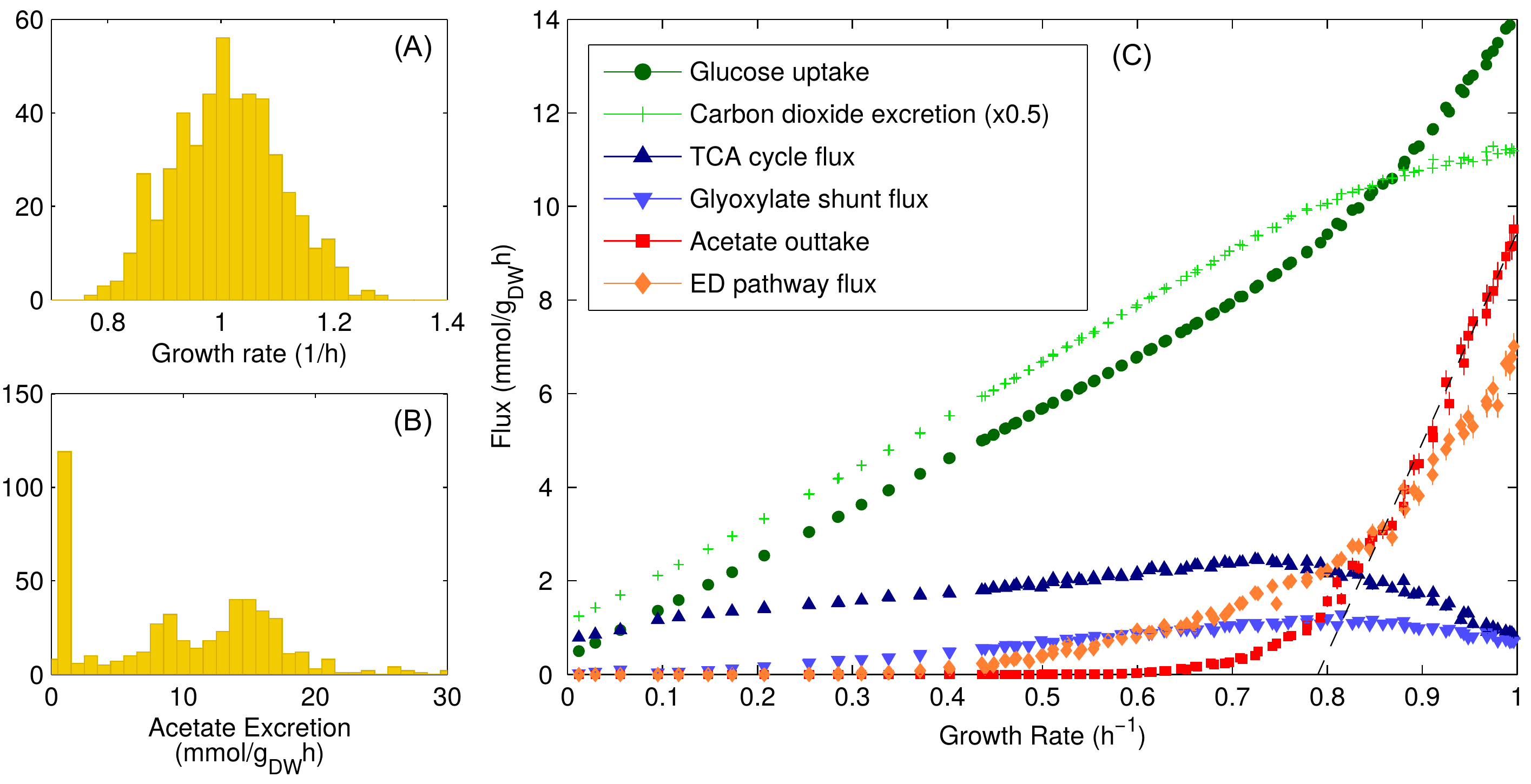}
\end{center}
\caption{\textbf{CAFBA solutions for {\it {\textbf{E. coli}}} in the heterogeneous case in carbon limitation.} Fluxes in glucose minimal medium computed at fixed $w_C\geq 0$ for different realizations of the E-sector weights, using $\langle w\rangle=8.8\times 10^{-4}$ gh/mmol and $w_{\max}/w_{\min}=10$ ($\delta=1$). (A) Histogram of the growth rates obtained from 1000 CAFBA solutions obtained using different randomly drawn weights for reactions in the E-sector and $w_C=0$. $\lambda$ peaks around $\lambda_{\max}=1/$h. (B) Histogram of the acetate secretion rates in the same conditions. Two classes of solutions are clearly visible, with roughly 25\% of states excreting less than 0.5 mmol/g$_{\DW}$h of acetate. The average secretion flux is close to 10 mmol/g$_{\DW}$h. (C) Average fluxes (glucose intake, carbon excretion, TCA, glyoxylate shunt, acetate excretion and ED pathway) versus the average growth rate. Each point represents the average of 1000 CAFBA solutions obtained with the same $w_C$ and different realizations of the weights of reactions in the E-sector. Both $x$ and $y$ error bars are shown. Different points are obtained by using different $w_C$ values. Acetate secretion is approximately linear at large values of $\lambda$. A line $v_{\ac}=s\times(\lambda-\lambda_{\ac})$ with $s=45$ mmol/g$_{\DW}$ and $\lambda_{\ac}=0.79/$h is shown for comparison.}
\end{figure}

Despite the crude approximations, CAFBA solutions appear to reproduce experimental findings with surprising accuracy. Fig 3A shows how the average acetate excretion rate compares with data from different experiments \cite{29,48,49,50,51}. Secretion rates from experiments using the MG1655 strain are consistent among each other (open triangles), as are results obtained with the NCM3722 and ML308 strains (open circles). CAFBA predictions are shown as solid circles for the two classes of strains. Data obtained with NCM3722 and ML308 were compared with CAFBA solutions obtained by setting $\lambda_{\max}=1/$h and hence $\langle w\rangle=8.8\times 10^{-4}$ gh/mmol. Instead, based on experimental evidence suggesting that MG1655 cells grow about 1/3 slower than the other two strains (see Fig D in the Supporting Text), for MG1655 we set $\lambda_{\max}=0.67$/h, leading to $\langle w\rangle=1.55\times 10^{-3}$ gh/mmol. With this choice, CAFBA quantitatively reproduces the growth-rate dependence of acetate secretion. Growth yields, instead, are less consistent across different experiments and/or strains, see Fig 3B. Without any further parameter tuning, CAFBA solutions capture the growth yields for NCM3722 and MG1655 at a quantitative level, although they fail for ML308. It should be noted that the differences in  yield among experiments done on the same strain (MG1655) suggest that other factors beyond the scope of this simple model might be at play, such as differences in growth conditions and/or maintenance requirements. 

\begin{figure}
\begin{center}
\includegraphics[width=15cm]{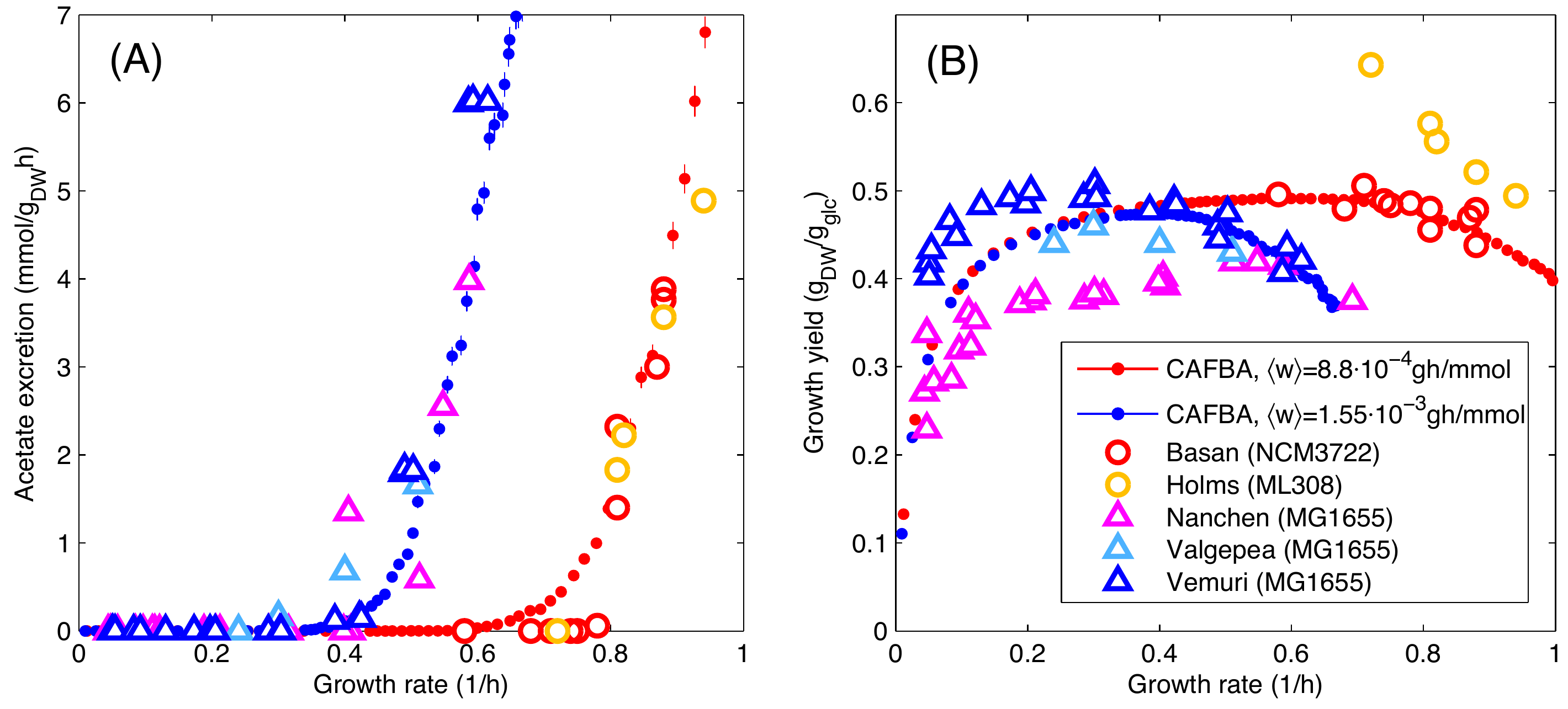}
\end{center}
\caption{\textbf{Comparison between CAFBA predictions and experimental data.} (A) Acetate secretion rates for {\it {\it E. coli}} cells grown in minimal glucose media, with data obtained from different datasets \cite{29, 48,49,50,51}. Full dots represent average CAFBA solutions (heterogeneous case) obtained with different degrees of carbon limitation (different $w_C$, averages over 500 solutions). Results were obtained with two different values for the average E-sector weight, namely $\langle w\rangle=8.8\times 10^{-4}$ gh/mmol (red) and $\langle w\rangle=1.55\times 10^{-3}$ gh/mmol  (blue). These choices  reproduce the acetate secretion rates of NCM3722 and ML308 (open circles) and MG1655 (open triangles) strains, respectively. (B) Same as panel (A), but for the growth yield. CAFBA predictions (red and blue filled circles) are obtained by averaging the ratio of the growth rate to the glucose intake flux, divided by the  molecular weight of glucose $\mu_{glc}=0.18$ g/mmol. Data points from \cite{29} have been converted using 1 mM/OD$_{600}/h=2$ mmol/g$_{\DW}$h. x- and y-error bars for the average CAFBA solutions are too small to be visible.}
\end{figure}

We have also analyzed how the flux patterns of various intracellular pathways are modulated by the growth rate. Results for the central carbon pathways are summarized in Fig E in the Supporting Text, with the fluxes through the TCA cycle and glyoxylate shunt consistently increasing in proportion as glucose is limited. A similar behavior has been observed in  measured expression levels of the corresponding enzymes \cite{8,51}. Glycolytic fluxes are heterogeneously regulated, due to the interplay between the EMP pathway, the ED pathway and the switch between glyoxylate shunt and the phosphoenolpyruvate carboxylase (PPC) reaction. The redox balance of the cell appears to be affected, as described in Fig F in the Supporting Text. Indeed we find that NADP transhydrogenase switches on at high growth rates, oxidizing NADH and reducing NADP$^+$, in agreement with the different roles of the two transhydrogenases, UdhA and PntA, as quantified by transcription data [52]. Moreover we observe a switch between two separate ubiquinol oxidase reactions, characterized by different abilities to generate proton-motive force, in agreement with studies focused on the crowding of the cell's membrane \cite{33}.

\subsection*{\textsf{Patterns of average flux for different carbon sources}}

We further tested CAFBA's ability to describe {\it {\it E. coli}} growth on carbon sources other than glucose. For illustration purposes, for each carbon source studied we have varied $w_C$ from zero to high values, so as to produce result in the entire range of growth rates $0\leq\lambda\leq\lambda_{\max}$, even though growth rates measured on individual carbon sources are always smaller than $\lambda_{\max}$ due to non-zero values of $w_{C,0}$ (see Table B in the Supporting Text). 

The typical behaviour of CAFBA solutions with different glycolytic carbon sources is remarkably consistent (see Fig 4). For each of the nutrients we tested, as the carbon supply becomes limiting, acetate excretion (Fig 4A) decreases almost linearly with growth rate, extrapolating to zero roughly at $\lambda_{\ac}\simeq 0.79/$h (continuous black line). By contrast, fluxes through TCA and glyoxylate shunt (Fig 4B and 4C) rise linearly with decreasing growth rate at fast growth, reaching a maximum close to $\lambda_{\ac}$ before decreasing at slower growth. The secretion rate of CO$_2$ (Fig 4D) almost always diminishes as $\lambda$ is reduced. For $\lambda<\lambda_{\ac}$ the decrease is linear, while it is non-linear for $\lambda>\lambda_{\ac}$. Altogether, for all carbon sources, results point to two distinct types of behaviors arising, respectively, below and above $\lambda_{\ac}$.

\begin{figure}
\begin{center}
\includegraphics[width=15cm]{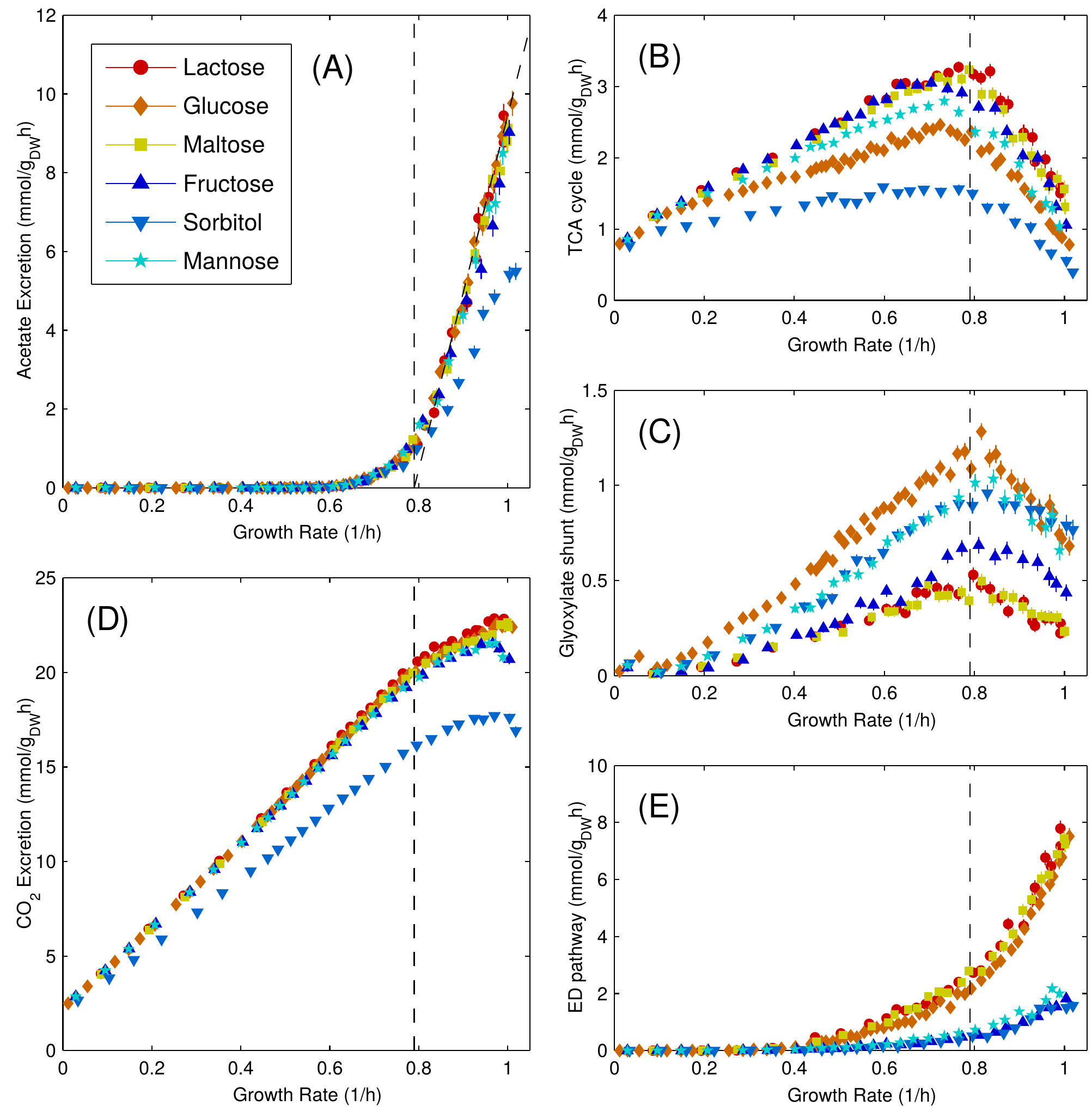}
\end{center}
\caption{\textbf{Average CAFBA solutions (heterogeneous case) for six different glycolytic carbon sources.} Each plot shows a different average flux, specifically: (A) Acetate secretion, (B) TCA cycle flux (represented by $\alpha$KG dehydrogenase), (C) Glyoxylate shunt flux (malate synthase), (D) CO2 secretion, (E) ED pathway flux, (6-phosphogluconate dehydratase). Each point represents the average over 500 solutions obtained with the same $w_C\geq 0$ and $\langle w\rangle=8.8\times 10^{-4}$ gh/mmol. Vertical lines at $\lambda_{\ac}=0.79/$h are shown for clarity. In panel (A), acetate secretion can be approximated, for $\lambda\geq\lambda_{\ac}$, with a straight line $v_{\ac}=s\times(\lambda-\lambda_{\ac})$ with $s=45$ mmol/g$_{\DW}$.}
\end{figure}

The Entner-Doudoroff (ED) pathway, an alternative to the Embden-Meyerhoff-Parnass (EMP) pathway, is used in {\it {\it E. coli}} for glucose catabolism at high growth rates \cite{53,54}. CAFBA solutions reproduce this feature, relying on the ED pathway from medium ($\lambda\simeq 0.3$/h) to high growth rates as shown in  Fig 4E. Interestingly, average fluxes are consistent for lactose, glucose and maltose on the one hand, and for fructose, sorbitol and mannose on the other. The reason is that, in the former group of substrates, the carbon source enters glycolysis as glucose-6P, which can be processed either by upper glycolysis or by the ED and pentose phosphate pathways. In the latter group, instead, carbon is transformed into fructose-6P, which is more conveniently processed into fructose biphosphate. A similar behavior is observed for phosphatated carbon sources or other substrates of the glycolytic or pentose pathways, see Fig G in the Supporting Text. The ED pathway, despite having a smaller ATP yield, requires a much smaller number of enzymes than the EMP pathway. Therefore, the use of ED over EMP may be the result of a proteome-saving strategy. Our findings thus agree with the conclusions of \cite{42,54,55}. The switch between the EMP and ED pathways sets in at a growth rate close to 0.3/h, well below $\lambda_{\ac}$, suggesting that it is independent of acetate secretion. Nonetheless, both features appear in CAFBA in order to cope with increasingly expensive proteins, in agreement with quantitative proteomics data \cite{46}.

On the other hand, CAFBA shows that a variety of strategies exist for cells growing on carbon substrates belonging to the lower part of glycolysis or to the TCA cycle, see Fig H in the Supporting Text. What these strategies share is an increased production of CO$_2$ at faster growth, and a vanishing activity of the ED pathway. The latter is of course due to the intrinsic glycolytic, as opposed to gluconeogenic, nature of the ED pathway.

\subsection*{\textsf{Comparison between CAFBA and FBA solutions}}

Standard FBA optimizes the growth yield subject to constraining the carbon intake flux. It is useful to compare CAFBA solutions with solutions obtained by FBA at the same growth rate and with glucose as the sole carbon source for both models. To do so, we have first solved Parsimonius Enzyme Usage FBA (pFBA, see \cite{73}) varying the bounds on glucose intake so as to obtain FBA solutions as a function of the growth rate. We shall denote them as $\mathbf{z}(\lambda)=\{z_i(\lambda)\}$. CAFBA solutions found upon varying $w_C$ lead instead to $w_C$-dependent mean growth rates $\overline{\lambda}(w_C)$. We shall denote CAFBA solutions obtained for a value of $w_C$ such that $\overline{\lambda}(w_C)=\lambda$ by $\mathbf{v}(\lambda)=\{v_i(\lambda)\}$. We have then computed, for a given set $\mathcal{R}$ of reactions of interest, the similarity index $q_{\mathcal{R}}$ called ``overlap'' and defined as \cite{mart}
\begin{equation}\label{fourteen}
q_{\mathcal{R}}(\lambda)=\frac{1}{N_{\mathcal{R}}}\sum_{i\in\mathcal{R}}\left\langle\frac{2 v_i(\lambda) z_i(\lambda)}{v_i(\lambda)^2+z_i(\lambda)^2}\right\rangle~~,
\end{equation}
where the sum is restricted to reactions in $\mathcal{R}$ and the brackets $\langle\cdots\rangle$ denote an average over 1000 different CAFBA solutions $\mathbf{v}(\lambda)$. If in each solution $v_i=z_i$ for each $i\in\mathcal{R}$, then $q_{\mathcal{R}}=1$. Conversely, the more the two flux vectors differ, the smaller $q$ gets. In particular, if in each solution $v_i=-z_i$ for each $i\in\mathcal{R}$, one finds $q_{\mathcal{R}}=-1$. Fig 5A shows the behavior of $q_{\mathcal{R}}$ versus $\lambda$ for different choices of $\mathcal{R}$. When all reactions are accounted for, $q$ is generally very large at low growth rates and decreases slowly as $\lambda$ increases. When focusing on individual pathways, one sees that the overlap for TCA fluxes (cyan) drops abruptly above the acetate onset point $\lambda_{\ac}\simeq 0.79/$h, where the growth yield of CAFBA solutions starts to reduce significantly compared to that of FBA solutions (Fig 5B, shown in red and blue symbols respectively). The overlap of fluxes in the glycolytic pathway instead diminishes with $\lambda$ in a more gradual way, corresponding to the smooth increase in the activity of the ED pathway, see Fig 4E. Thus, as the growth rate increases, CAFBA solutions cross over from flux distributions that maximize the growth yield (slow growth) to a regime in which low-yield fermentation, accompanied by carbon overflow and energy spilling, is favored (fast growth).

\begin{figure}
\begin{center}
\includegraphics[width=15cm]{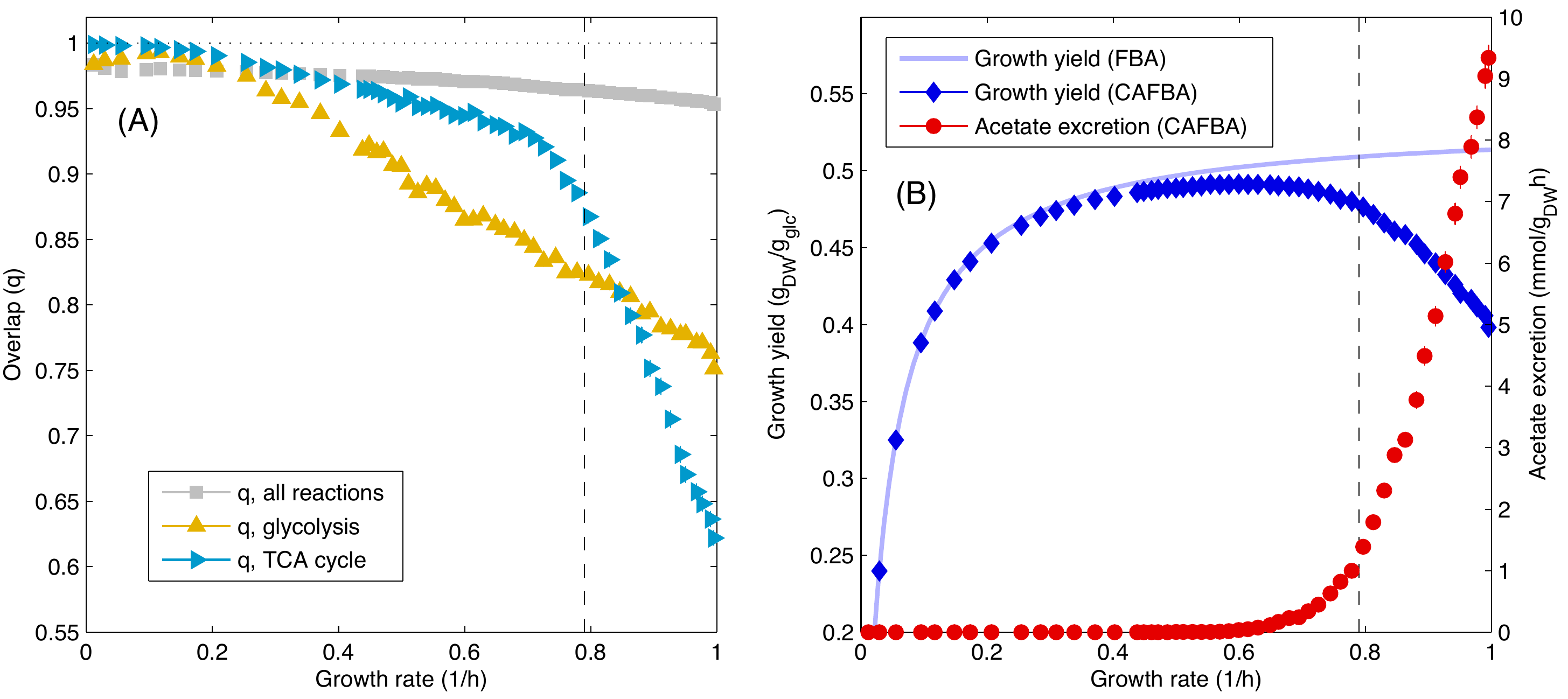}
\end{center}
\caption{\textbf{Overlap between FBA and CAFBA solutions.} (A) Overlap $q$ between pFBA and CAFBA solutions as a function of growth rate, computed for three different reaction sets: all  reactions included in the reconstruction, reactions in the glycolysis/gluconeogenesis pathway, and reactions in the TCA cycle. FBA fluxes were computed for the same glucose influx as the CAFBA solution and then interpolated at the growth rates of CAFBA solutions in order to plot the overlap as a function of $\lambda$. (B) Growth yield and acetate secretion from CAFBA, together with the FBA-predicted growth yield. In both panels the value $\lambda_{\ac}=0.79/$h is marked by a vertical dashed line.}
\end{figure}

\subsection*{\textsf{Growth rate-dependent biomass composition}}

In CBMs, the energetic cost of anabolic pathways is accounted for by the stoichiometry of the network. By contrast, the energetic requirements of growth (e.g. protein synthesis) and homeostasis must be included separately as an additional ATP hydrolysis flux $v_{\ATP}$. In metabolic models, the latter is assumed to be linearly related to the growth rate, i.e. $v_{\ATP}=v_{\ATPM}+\beta_{\ATP}\lambda$ \cite{15}. The first term is a growth-rate independent maintenance flux that represents the energy required to sustain basal cellular activities. The second term, instead, accounts for $\lambda$-dependence through a coefficient $\beta_{\ATP}$ that fixes the moles of ATP to be hydrolyzed per gram of dry weight. The values of $v_{\ATPM}$ and $\beta_{\ATP}$ are usually fitted from growth yield curves \cite{16}, and different metabolic reconstructions of {\it {\it E. coli}} use different numerical values for both of them, see \cite{45,56,57} and Table C in the Supporting Text. However, as the cell's composition (and specifically the amounts of RNA, DNA, proteins, fatty acids, etc.) adjusts with the growth rate, biomass coefficients, including the demand of growth-related ATP, are in general $\lambda$-dependent \cite{39}. A natural question to ask at this point is how cellular ATP requirements impact the shift between respiration and fermentation.

Results obtained by solving CAFBA with $\lambda$-dependent biomass composition are shown in Fig 6 (open symbols), together with the solution obtained for constant biomass composition at the same $\langle w\rangle=8.8\times 10^{-4}$ gh/mmol (filled blue circles). We tested CAFBA predictions with three different values of $\beta_{\ATP}$ while keeping $v_{\ATPM}$ fixed: (i) the default value for iJR904 model, (ii) the default value for the iAF1260 model, which is 30\% larger than (i) \cite{56}, and (iii) a value 30\% smaller than (i). One sees that, for the same ATP hydrolysis parameter (open and filled blue symbols), solutions for the two versions of CAFBA nearly overlap. On the other hand, both the slope and the onset growth rate $\lambda_{\ac}$ for acetate \red{secretion} appear to depend on the value of $\beta_{\ATP}$. Likewise,the flux through TCA increases with $\beta_{\ATP}$ so as to satisfy energetic requirements. The growth yield and maximum growth rate $\lambda_{\max}$ obtained at $w_C=0$ decrease accordingly. However, if we tune $\langle w\rangle$ to fix $\lambda_{\max}=1$/h for each value of $\beta_{\ATP}$, acetate secretion starts consistently at $\lambda_{\ac}\simeq 0.8$/h (Fig J in the Supporting Text), implying that energetic costs do not affect the ratio $\lambda_{\ac}/\lambda_{\max}$.

\begin{figure}
\begin{center}
\includegraphics[width=15cm]{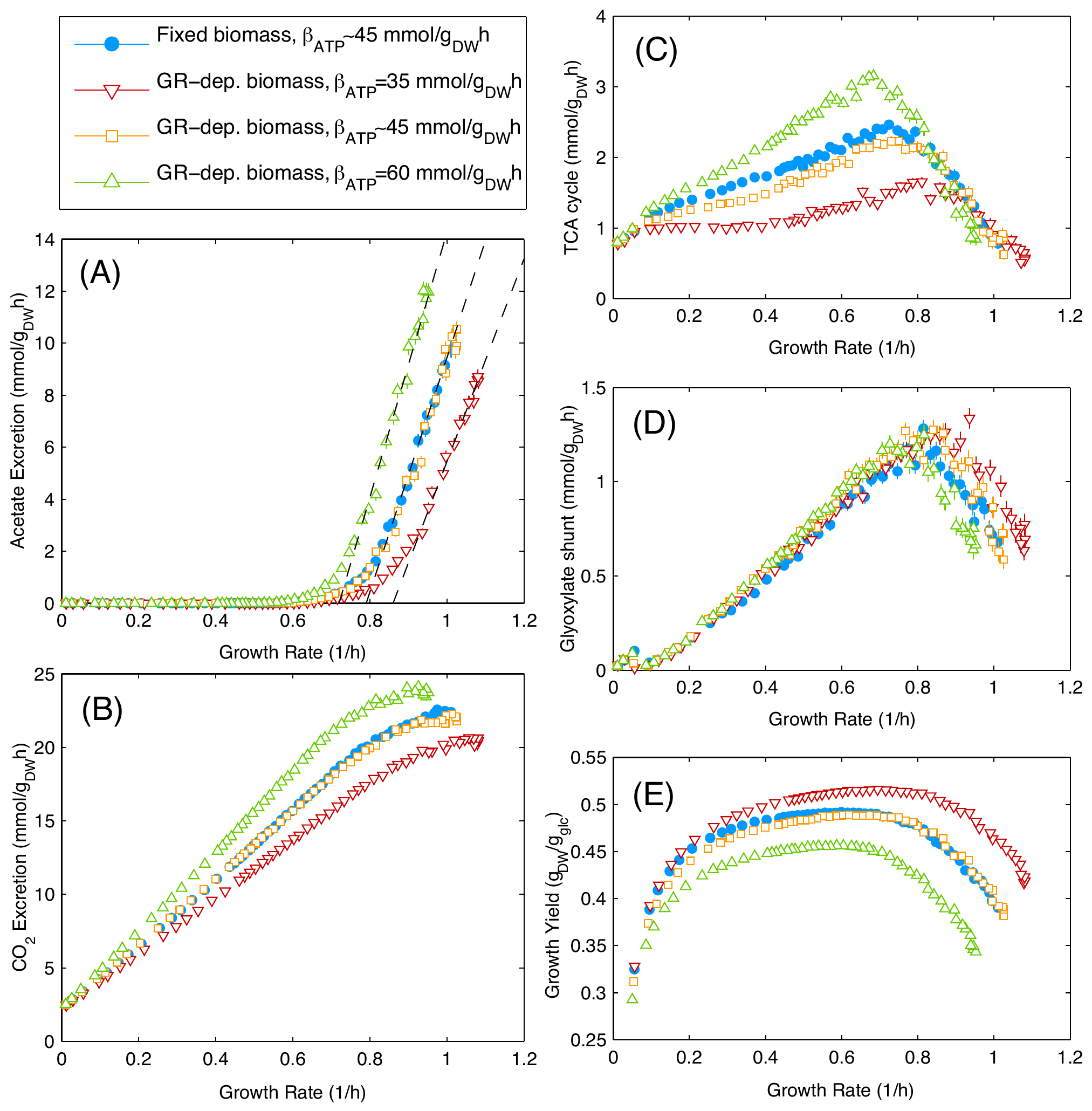}
\end{center}
\caption{\textbf{CAFBA solutions with growth rate-dependent biomass composition.} Representative fluxes obtained by CAFBA for {\it E. coli} growth in glucose minimal medium with fixed (blue points) and variable biomass composition (in open red, yellow and green markers for three different values of the $\lambda$-dependent ATP hydrolysis rate $\beta_{\ATP}$). (A) Acetate secretion rate, (B) CO$_2$ secretion rate, (C) flux through TCA cycle ($\alpha$KG dehydrogenase), (D) flux through glyoxylate shunt (Malate synthase), (E) growth yield. No significant differences are observed between the constant and $\lambda$-dependent cases for $\beta_{\ATP}=45.5608$ mmol$_{\ATP}$/g$_{\DW}$, corresponding to the default value for the iJR904 model. We also show, for comparison, results obtained for larger and smaller values of $\beta_{\ATP}$. The acetate secretion rate can always be fitted by a linear function of $\lambda$, i.e. $v_{\ac}=s\times(\lambda-\lambda_{\ac})$, albeit with different slopes and intercepts. The three dashed lines correspond to $s= 39, 45 , 51$ mmol/g$_{\DW}$, respectively, while $\lambda_{\ac}= 0.86, 0.79 , 0.72/$h, respectively. We also indicate $\lambda_{\ac}=0.79/$h with a vertical dashed line in all panels. In all cases we set $\langle w\rangle=8.8\times 10^{-4}$ gh/mmol, $w_C\geq 0$ and $w_{\max}/w_{\min}=10$.}
\end{figure}

\section*{\textsf{Discussion}}

In this work we have introduced CAFBA, an extension of FBA inspired by the proteome allocation scenario underpinned by bacterial growth laws \cite{3,5,21,58}. By integrating a single additional global constraint in FBA, CAFBA formulates the interplay of growth and expression in metabolism as a simple and elegant growth-rate optimization problem, with the same computational complexity as standard FBA. States of optimal growth found by CAFBA therefore encode for optimality from both an energetic and a proteome allocation perspective. A most distinctive feature of CAFBA lies in the extremely simple empirical inputs required to make quantitatively accurate predictions. The 3 parameters on which the proteome allocation constraint relies, namely $w_{R,0}$, $\phi_{\max}$ and $\lambda_{\max}$, can be easily obtained in experiments \cite{5,6}. All other parameters can be set based on these 3 numbers. 

CAFBA predictions obtained for {\it E. coli} by averaging solutions over protein  costs are found to be close to growth-yield maximizing solutions at slow growth. As growth gets faster, a continuous switch to a regime characterized by carbon overflow occurs. The onset of carbon overflow (at a growth rate denoted as $\lambda_{\ac}$) turns out to be largely independent of the nature of the glycolytic substrate. The ratio $\lambda_{\ac}/\lambda_{\max}$, with $\lambda_{\max}$ the fastest achievable growth rate, is indeed a remarkably robust quantity, that is roughly independent of the empirical parameters that characterize the proteome allocation constraint. Rather, it is mainly influenced by the weights of the biosynthetic reactions, $\{w_i\}$. These results strongly support the picture according to which acetate secretion is part of an optimal strategy to cope with increasing protein costs at high growth rates \cite{7,8, 29,46}. 

CAFBA easily allows to model cellular metabolic activity in a variety of conditions, including translational limitation \cite{8, 22} and protein overexpression \cite{5}. As more growth laws are being characterized in different organisms \cite{66,67,68}, CAFBA's application range is likely to expand significantly. Note C in the Supporting Text details how to port CAFBA to growth conditions and/or bacterial species different from those considered here. Other growth-maximizing  organisms may also be studied by CAFBA if the required ingredients (network structure, biomass composition, empirical inputs) are available. Going beyond cell-autonomous models, CAFBA may prove highly effective for characterizing trophic interactions (e.g. cross-feeding) in microbial communities by treating excreted metabolites as potential nutrients \cite{69}. CAFBA therefore provides a conceptually simple and computationally efficient platform that can be easily adapted and calibrated to describe the metabolism and growth of different organisms, making it a versatile tool for the computational modelling of interacting species in complex environments.

\subsection*{\textsf{Comparison of CAFBA to other models}}

Under carbon limitation, solutions of CAFBA are obtained by varying the parameter $w_C$, a proxy for the extracellular carbon level representing the proteome cost of the C-sector (carbon intake). In essence, for any given substrate level, CAFBA allocates the C-sector proteins per unit flux by simultaneously optimizing the allocation of the proteome fractions required to sustain biosynthesis and translation in order to maximize growth. The use of $w_C$ as a control parameter as opposed to directly dialing the nutrient intake flux is one of the elements that distinguish CAFBA from closely related CBMs like FBA \cite{15}, FBAwMC \cite{31, 32} and ME-models \cite{19, 20}. In fact, the CAFBA constraint effectively reduces to a finite capacity constraint similar to the one that characterizes FBAwMC when an upper bound on the glucose intake flux is used to modulate growth at fixed $w_C=0$ (see Note B in the Supporting Text). (Note however that tuning nutrient levels as opposed to nutrient influx was employed in RBA to model the substitution between low affinity and high affinity cysteine transporters in {\it B. subtilis} \cite{18}.) 

Secondly, CAFBA does not provide the detailed mechanistic description of gene expression and protein synthesis conveyed by ME-models and RBA, whose definition includes, for instance,  explicit variables for macromolecular concentrations (ribosomes, DNA, RNA, etc.). Rather, it relies on an effective formulation based on empirical growth laws and (when desired) on a growth-rate dependent biomass composition. In this light, while less comprehensive than its closely related CBMs, the CAFBA scheme highlights the key biological ingredients constraining proteome allocation.

On a more technical level, both RBA and ME-models are intrinsically non-linear and handle non-linearity by approximating their underlying optimization problems through sequences of linear feasibility problems. In CAFBA, even the worst case is solved through a fast iterative algorithm involving a small number of LP problems.

Finally, the optimal proteome allocation problem posed by CAFBA can be seen as an assumption of ``optimal enzymatic efficiency'', close to that underlying FBA approaches based on flux minimization \cite{holz,cakir}.

\subsection*{\textsf{Choice of parameters}}

One of the strong points of standard FBA consists in its reliance on the stoichiometric matrix and on thermodynamic reversibility constraints alone, making kinetic parameters unnecessary. CAFBA's proteome allocation constraint in principle introduces a large number of additional parameters related to reaction and/or transport kinetics that, for the most part, are either uncharacterized or inferred from in vitro studies performed in different biochemical conditions \cite{59}. This raises the issue of parameter selection. 

Two of the constants entering the proteome constraints (namely $w_R$ and $\phi_{\max}$) are obtained directly from empirical growth laws. With $w_C$ acting as the control variable, the only free parameters left are the weights $w_i$ characterizing intracellular reactions. While quantitative CAFBA predictions appear to be dependent on their specific values, the qualitative behaviour of the solutions is not. Furthermore, the scenario obtained by averaging CAFBA solutions over different choices of the $w_i$ quantitatively reproduces experimental findings for acetate secretion and growth yield. These results point to a considerable degree of robustness of the CAFBA framework against fluctuations in parameter values. Notice however that the CAFBA picture can be further improved upon tuning the weights of individual reactions. For example, by increasing the average weight of reactions involved in respiration one sees a shift in the onset of acetate secretion and the value of $\lambda_{\ac}/\lambda_{\max}$ changes, see Fig K in the Supporting Text. On the other hand, parameters can also be tuned according to empirical evidence so as to allow for a more thorough comparison with experiments performed on different strains and/or growth conditions, e.g. concerning intracellular fluxes (see Fig L in the Supporting Text). 

In perspective, detailed flux measurements may allow to estimate typical weights for each pathway, and possibly even for individual enzymes, opening for the possibility to better calibrate the model and obtain completely quantitative predictions. Our work here has aimed at keeping the number of parameters as small as possible. In this light, many emerging features of the interplay between metabolism and gene expression appear to be mostly determined by the topology of the metabolic network. Elucidating the origin of this simplification is a foremost theoretical challenge for future studies of metabolic systems.

\subsection*{\textsf{On using average fluxes}}

The need to resort to an averaging procedure in order to reproduce bulk measurements for the growth yield and the acetate excretion forces us to ask whether the CAFBA averaging may have some further meaning. We offer here two possible scenarios.

The first one is based on the fact that, even in well controlled growth conditions, cells in a population are normally heterogeneous, as transcription levels, protein abundances, reaction fluxes and instantaneous growth rates may change significantly from one to the other \cite{60,61,62}. This would in turn reflect in fluctuations in the values of each $w_i$ across cells. Averaging over different choices for the weights could then simply be interpreted as averaging over a population of heterogeneous cells (as would seem appropriate in modelling batch culture or chemostats). 

The alternative scenario presupposes that, even in absence of any cell-to-cell variability, cells may not be able to perfectly adjust fluxes to the distribution maximizing the instantaneous growth rate. This may occur for different reasons. First, the regulatory machinery needed to perform protein allocation requires by itself an investment of metabolic and proteomic  resources \cite{63,64,65}. This burden becomes more severe as the regulatory system gets more sensitive and fine-tuned, and, clearly, CAFBA does not account for it. Secondly, environments where cells grow are always fluctuating. Any regulatory machinery implementing fast adjustments in response to small environmental changes will necessarily come at a cost that will negatively affect the growth rate. Under such constraints, regulatory programs selected over evolutionary time scales may prefer to maximize an {\it average} growth rate, the average being taken over life process history. The actual regulatory programs implemented would then balance the trade-off between the costs of not being exactly in the instantaneously optimal growth state and the costs of adjusting regulation too frequently in ``natural'' conditions (not those provided in the laboratory). An interpretation of the CAFBA averaging prescription would then be that it is a way to implement an ``average'' strategy that smooths the output upon variations of the environmental conditions.

In both of the above scenarios, CAFBA points to the emergence of acetate excretion as triggered by regulatory system(s) sensing the abundance of the carbon source and the balance of biomass synthesis and energy generation \cite{29}. We note that carbon overflow in {\it E. coli} has been proposed to be modulated by catabolite repression mediated ACS down-regulation \cite{49}. Discriminating between the two scenarios we have just presented could be achieved already in bulk experiments, by changing the weight of specific enzymes over time (e.g., by expressing useless proteins specific to certain pathways) and monitoring whether the associated fluxes adjust dynamically in real time. Naturally, tests of cell-to-cell heterogeneities would also allow to favor one scenario over the other.

Finally, we address the magnitude of fluctuations of the weights $w_i$. The spread in the enzyme catalytic rates $k_{\cat,i}$, as tabulated in databases, is notoriously broad, exceeding 3 orders of magnitude \cite{32,59}. In the absence of more refined information, it is reasonable to expect that the weights $w_i$ should fluctuate by about the same amount. However, we have seen that unrealistic results are generated by CAFBA if weights are allowed to fluctuate more than 10-fold. Therefore, either the true width of the distribution of the weights $w_i$ is much smaller than what is suggested by the values of $k_{\cat,i}$ estimated in vitro, or weights are subtly distributed across pathways in such a way that strong compensatory effects occur that reduce fluctuations. With steady improvements in proteomic methods, it may soon be possible to quantitatively determine these parameters empirically and elucidate this puzzle.

\section*{\textsf{Materials and Methods}}

\subsection*{\textsf{Optimization problem}}

Given a metabolic network encoded by a stoichiometric matrix $\mathbf{S}=\{S_{\mu i}\}$, CAFBA is stated in the case of carbon limitation as
\begin{eqnarray}
 \max_{\mathbf{v}} \lambda \quad\text{subject to} & \text{(i)} & ~~~\sum_i S_{\mu i}v_i=0 ~~~\forall \mu~  \label{eq:SI_CAFBA_1} \\
						   & \text{(ii)} & ~~~ \ell_i \le v_i \le u_i ~~~\forall i~  \label{eq:SI_CAFBA_2}   \\
						   & \text{(iii)} & ~~~ w_C v_C + \sum_i w_i |v_i| + w_R \lambda = \phi_{\max} ~, \label{eq:SI_CAFBA}
\end{eqnarray}
where $\lambda$ denotes the growth rate, $\mathbf{v}=\{v_i\}$ is a flux vector, and $(\ell_i,u_i)$ represent lower and upper bounds for each flux $v_i$, respectively. Condition \eqref{eq:SI_CAFBA} corresponds to the proteome allocation constraint $\phi_C+\phi_E+\phi_R+\phi_Q=1$, with $v_C\geq 0$ being the active glucose intake flux and with the sum in $\sum_i w_i |v_i|$ running over all enzyme--catalyzed reactions except for transports, exchanges and carbon intake pathways. The biomass flux $\lambda$ and ATP maintenance reaction are also excluded from (\ref{eq:SI_CAFBA}). 

In principle, CAFBA is a MILP (Mixed Integer-Linear Programming) problem due to the  presence of absolute values in \eqref{eq:SI_CAFBA}. However, in CAFBA they can be disposed of by splitting each flux $v_i$ into a forward $v^+_i$ and a backward $v^-_i$ component, both non-negative. Note that if either $v^+_i$ or $v^-_i$ can be set to zero for each $i$, net fluxes $v_i=v_i^+-v_i^-$ are univocally determined, one has $|v_i|=v_i^++v_i^-$ for absolute values, and CAFBA becomes equivalent to \begin{eqnarray}
 \max_{\mathbf{v^+},\mathbf{v^-}} \lambda \qquad\text{subject to} & \text{(i)} & \sum_i S_{\mu i}(v_i^+-v_i^-)=0 ~~~\forall \mu~  \label{eq:SI_CAFBA_1b} \\
						   &\text{(ii)} & 0 \le v_i^- \le -\ell_i \qquad 0 \le v_i^+ \le u_i ~~~\forall i~  \label{eq:SI_CAFBA_2b}   \\
						   & \text{(iii)} &w_C v_C + \sum_i w_i (v_i^++v_i^-)+ w_R \lambda = \phi_{\max} ~, \label{eq:SI_CAFBA_b}
\end{eqnarray}
which is a simple LP problem rather than a MILP. The key observation is that, as long as $\lambda$ is maximized, CAFBA actually adjusts fluxes so that either the forward or the backward component vanish for each $i$. Indeed, a necessary condition for maximizing $\lambda$ is that the quantity $v_i^++v_i^-$ is minimized for each $i$, which, at fixed $v_i=v_i^+-v_i^-$, is achieved by setting either $v_i^+$ or $v_i^-$ to zero.  Therefore CAFBA reduces from a MILP to a LP problem. 

Note that, because of the tight link between CAFBA and flux minimization, degeneracies in CAFBA solutions can only arise from the presence of (a) futile loops or (b) pathways that perform the same overall chemical conversion with the same flux at the same proteome cost, and which therefore can be used alternatively. In CAFBA with heterogeneous weights, however, the chance that two equivalent flux configurations have exactly the same total weight is negligible, since weights are i.i.d. random variables. On the other hand, futile loops only concern transports that do not involve the main carbon source and therefore are not included explicitly in the CAFBA constraint. These loops however do not affect other fluxes  and are easily spotted and removed, either by manually shutting off redundant processes or by including them into the proteome allocation constraint with an arbitrarily small but non-zero weight. Therefore, each instance of the inhomogeneous CAFBA scheme has a unique solution almost surely. Because our main results are obtained in this framework, alternate optima are in practice not an issue in CAFBA.

\subsection*{\textsf{Implementation}}

We implemented CAFBA on the E.~coli iJR904 genome-scale model \cite{45}, comprehensive of 761 metabolites and 1075 reactions, as a Matlab$^\circledR$ function, using the COBRA Toolbox \cite{72} to load the network reconstruction with a minor modification. Specifically, we shut off the glucose dehydrogenase reaction, since it is only functional if the cofactor pyrroloquinoline quinone is supplied in the environment (see the Ecocyc \cite{71} entry on the enzyme). Both GLPK- and Gurobi-compatible CAFBA solvers for Matlab are provided as additional supplementary material, along with a small set of utility functions. Running times for a single CAFBA optimization of the iJR904 network with a common laptop (single thread of an Intel Core i7--2630QM CPU @ 2.00GHz) are around $0.12$ s for the GLPK (version 4.47) LP solver and $0.05$ s the Gurobi Optimizer (version 5.6) solver. For comparison, the time required to compute the standard FBA solution for the same network with the COBRA toolbox using GLPK is around $0.06$ s.

\subsection*{\textsf{Case of growth-rate dependent biomass}}

The fact that cells adapt their composition with the growth rate  \cite{2, 4, 5} implies that biomass composition is itself $\lambda$-dependent. Growth-rate dependent biomass coefficients (see e.g. \cite{39} for {\it {\it E. coli}}) indeed reflect empirical knowledge of how the amounts of RNA, DNA, proteins, fatty acids, etc. are modulated by $\lambda$. While constraint-based models such as FBA and CAFBA with growth-dependent biomass are non-linear, approximate solutions can be obtained efficiently by simple iterated LP protocols as follows: (a) starting from a given biomass vector, solve the model by optimizing the growth rate; (b) update the biomass composition to the computed optimal growth rate using the prescribed set of $\lambda$-dependent biomass coefficients; (c) iterate until a solution is reached, such that further iterations do not change the optimal growth rate within a desired precision. For CAFBA, this procedure typically converges in a very small number of iterations (see Fig I in the Supporting Text). Further details about the case of growth rate-dependent biomass composition and the iterative algorithm for computing the optimal FBA or CAFBA solutions are  given in Note F in the Supporting Text.

\subsection*{\textsf{Extension to different growth media and/or organisms}}

Besides the full study of the {\it {\it E. coli}} iJR904 model, we have tested CAFBA on the more recent reconstructions iAF1260 \cite{56} and iJO1366 \cite{57}, obtaining very similar results. COBRA-compatible Matlab functions \cite{72} to run CAFBA on these models are provided as additional supplementary material. Note C in the Supporting Text describes in detail how to port CAFBA to different growth media, nutrient limitations and/or bacterial species. However, provided the input coming from empirical growth laws is available together with the network structure and the biomass composition, the CAFBA framework can in principle be extended to growth-maximizing organisms other than bacteria.

\subsection*{\textsf{Acknowledgments}}

MM gratefully acknowledges the hospitality for his visit to Hwa labs at UCSD. TH gratefully acknowledges the hospitality of U. Paris-Sud and Sapienza, U. Roma for his visits. The authors acknowledge members of Hwa lab for contributing growth rate data for comparing between MG1655 and NCM3722 strains. Useful discussions with S. Hui, R. Mulet, A. Pagnani, A. Samal, A. Tramontano and Z. Yang are gratefully acknowledged.




%
%
%

\includepdf[pages=-]{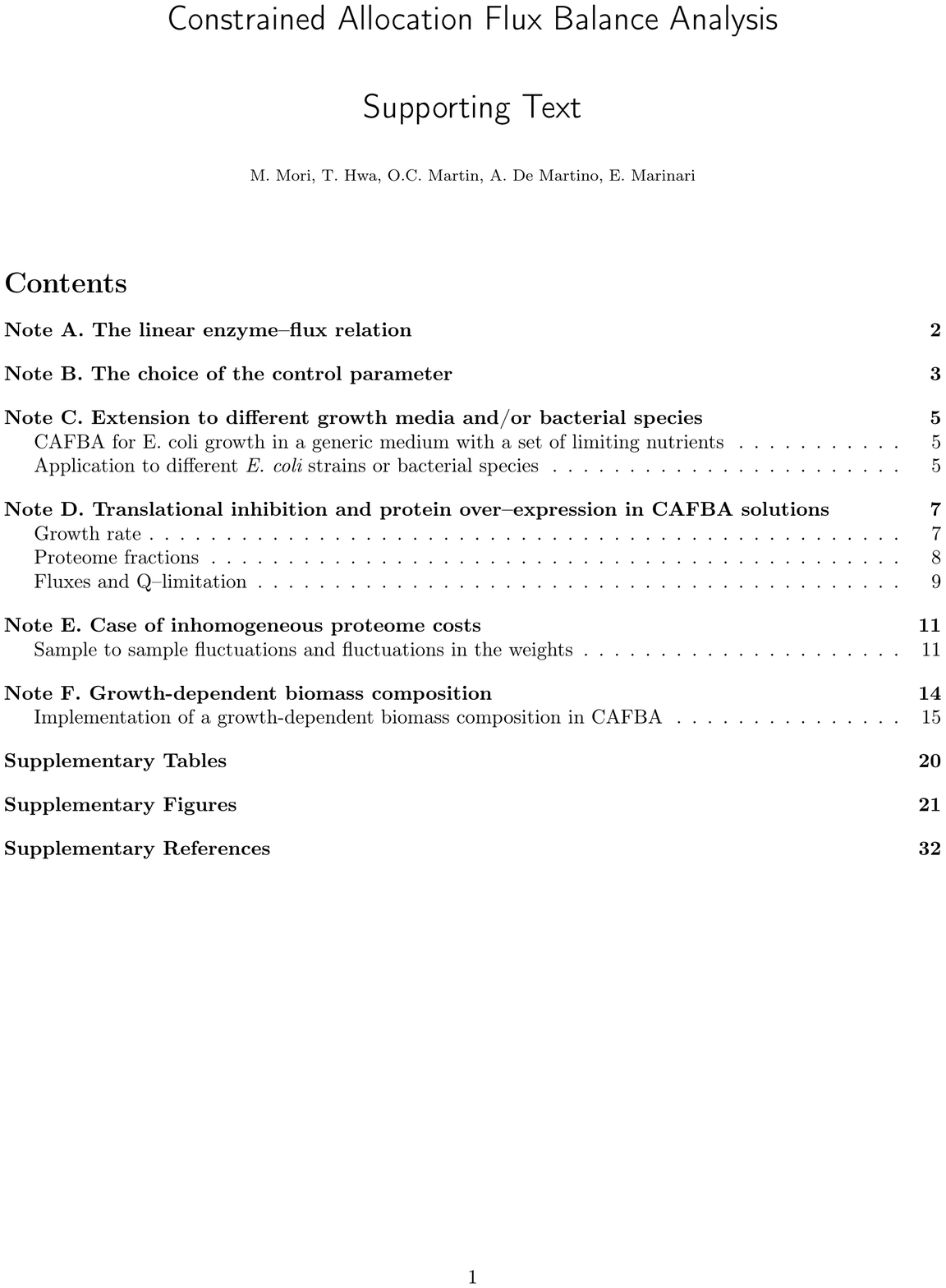}

\end{document}